\documentstyle[aps,floats,preprint]{revtex}
\begin{document} 

\title{ Stripe Dynamics, Global Phase Ordering and Quantum Criticality in High $T_c$ Superconductors
      }
\author{WenJun Zheng}
\address{ Department of Physics and Center for Material Research, Stanford University, Palo Alto, California 94305}
       
\date{\today}       
\maketitle 

\begin{abstract}
    By modeling the stripe phase in cuprates as spin gapped stripes coupled 
to the RVB liquid of preformed electron pairs, I derive the low energy effective theory of 
 the RVB phase variable. It is found that the effect of stripe dynamics
 (including both longitudinal and transverse modes) leads to incipient temporal phase stiffness in the RVB liquid, which tunes a quantum phase transition toward a superconducting ground state with global phase order.
 Physical consequences of this quantum criticality are discussed.

\end{abstract} 
\hspace{1cm}

\section{Introduction}

The relevance of quantum criticality to the mechanism of high $T_c$ superconductivity in cuprates
 has captured considerable interest in the theoretical 
community. One scenario \cite{SDW} argues that the proximity to a quantum critical point associated with anti-ferromagnetic(AFM) ordering is responsible for the anomalous normal state properties and the pairing mechanism that leads to d-wave superconductivity. There is also recipe that  emphasizes the competition between AFM and superconductivity(SC) orders \cite{zhang}, which is likely controlled 
by hidden quantum critical points \cite{laughlin}. 

  Recently there have been convincing experimental evidences that support the presence of stripe ordering (both dynamic and static) in typical cuprate materials such as $La_{2-\delta}Sr_\delta CuO_4$, $YBa_2Cu_3O_{7-\delta}$ and $Bi_2Sr_2CaCu_2O_{8+\delta}$ (Bi-2212) etc \cite{stripe-exp}. This offers new possibilities of quantum critical
 scenarios, considering that stripe phase requires charge order 
 to be locked to local AFM order in its competition with SC ordering. Motivated by 
these results, suggestions of new critical point associated with charge ordering \cite{italy} 
were advanced. 
   
   Among many unresolved issues on relations between various orders, the interplay between stripe order and SC order has been under hot discussions\cite{stripesc}. In the theory suggested in ref \cite{EKT}, it is argued that Cooper pairing is induced in hole-rich stripes through "spin proximity" effect caused by pair tunneling between stripe and insulating background, and global phase ordering occurs at a lower energy scale determined by the inter-stripe Josephson coupling which is enhanced by transverse zero-point fluctuations of stripes \cite{EK1}. In parallel to the above 
recipe, I recently suggested a scenario \cite{zheng} where stripes are 
coupled to an RVB ( Resonating Valence Bond) spin liquid background through single-particle hopping, which results in the generation of two quantitatively different gaps 
( normal state pseudo-gap and superconducting gap) by strong 
pairing correlation inherent in the RVB environment. 
In both of these scenarios, the dynamical stripes play the central role in 
accommodating the seeds of Copper pairs for the later establishment of superconductivity 
 order which results from the overlaps of the SC wave function between neighboring {\it one 
dimensional superconducting stripes}, while the hole deficient regions between stripes are less relevant and treated as being remnant of the undoped antiferromagnet with reduced 
local magnetization which could only compete with the  superconducting order. It is then natural to ask: Is it the only possible route toward high
 $T_c$ superconductivity in striped phase of cuprates? In this paper, we will attempt to 
 explore an alternative scenario based on the same microscopic model proposed in ref \cite{zheng}, albeit from a different viewpoint of the role of stripe dynamics, namely, the effect of dynamical stripes in turning a pre-paired RVB spin liquid into a superconductor.

  The relevance of RVB spin liquid to high $T_c$ superconductivity was suggested by Anderson more than a decade ago \cite{anderson}. The basic idea is that 
the undoped cuprates have a novel quantum disordered ground state which is 
the resonating superposition of different configurations of local singlet pairs
 (so called valence bonds). Upon hole doping, these localized electron pairs 
are gradually liberated and become Cooper pairs which condense into a superconducting ground state. Around this proposal, there have been a lot of discussions
 and controversies in the theoretical community, and it is still inconclusive \cite{laterRVB}.
 We note that beyond the detailed formulation of RVB theory, there is one important aspect which has enjoyed a broader acceptance ,
that is, strong local pairing correlation is 
present even in the normal state of under-doped cuprates although the global phase
 coherence is established only at a lower temperature. Despite the general argument
 by Emery and Kivelson \cite{EK2} about the effect of low carrier density on
 reducing phase stiffness, the concrete mechanism that is
responsible for turning a pre-paired but incoherent RVB liquid into a
superconductor with global phase coherence, is albeit unclear.
 It is interesting to return to this issue now, thanks to the development of 
experiments which provide precious information and constraints on any serious 
theoretical effort to understand the high $T_c$ mechanism, such as the 
presence of stripe correlation in the charge degree of freedom which must be 
 taken into account in discussing the above issue. 

 In this work, we will try to address this issue, 
based on the consideration of stripe
 dynamics  and its effect on phase ordering in the RVB spin liquid which then gives rise to a global superconducting order.     
The detailed formulation is provided in next section and the last section is devoted to discussions and conclusions.

\section{Formulations} 
Based on the two-component stripe picture suggested in ref \cite{zheng}, one can start with the total microscopic Hamiltonian as follows,
\begin{eqnarray}
              H(c,c^+,d,d^+)&=&H_{1D}(d,d^+) + H_{RVB}(c, c^+)   \\   \nonumber
   ~~~~~~~~~~~~                    &+&H_{couple}(c,c^+,d,d^+) ,
\end{eqnarray}
 where $c$, $c^+$ and $d$, $d^+$ represent the annihilation and creation operators of a single electron in 2D RVB background and 1D stripe, respectively. 
In the undoped background, there is on average one electron per site and 
the charge degree of freedom is frozen (which is however gradually mobilized when  the coupling with stripes develops with hole doping), and only spin exchange interaction is relevant at low energy scale, which is 
responsible for the singlet formation in RVB state. Within the stripes, where 
doped holes concentrate, both charge and spin degrees of freedom are active
 at low energy scale. 

Now, let us treat these three parts one by one as follows in order to obtain the low energy effective theory:

{\it{Hamiltonian of 2D RVB spin liquid $H_{RVB}$:}}
Since the undoped background is at half-filling, one can start with 
the 2D antiferromagnetic Heisenberg model and perform
a routine Hartree-Fork decoupling which leads to the  RVB Hamiltonian $H_{RVB}$ \cite{anderson},
\begin{eqnarray}
   H_{RVB}&=&J\sum_{<ij>}(S_i S_j-1/4) \\ \nonumber
          &=&-J\sum_{<ij>}b^+_{ij}b_{ij}  \\  \nonumber
&=&-J\sum_{<ij>}(\Delta_{ij}b^+_{ij}+h.c. -|\Delta_{ij}|^2),
\end{eqnarray}
where $b^+_{ij}=\frac{1}{\sqrt{2}}[c^+_{i,\uparrow}c^+_{j,\downarrow}-c^+_{i,\downarrow}c^+_{j,\uparrow}]$. 
Integrating over fermion variables, one can get the "free energy " of RVB order parameters(OP) $\Delta_{ij}$ \cite{anderson2}:
\begin{eqnarray}
 F_{RVB}&\approx& a \sum_{<ij>}|\Delta_{ij}|^2 + b\sum_{<ij>}|\Delta_{ij}|^4  \\ \nonumber
&+& c \sum_{plaquette[ijkl]}(\Delta_{ij}^*\Delta_{jk}\Delta_{kl}^* \Delta_{li} + h.c.)+..., 
\end{eqnarray}
where $a$,$b$,$c$ are parameters derived from microscopic model calculations.
 Then approaching the continuous limit by 
coarse graining:  $ \Psi(\vec{r} \leftarrow \frac{\vec r_i+\vec r_j}{2}) \leftarrow {\it~~local~~average~~of~~} |\Delta_{ij} |\exp (i \theta_{ij})  $, one can arrive at the following effective action
\begin{eqnarray}
\label{EQ0}
    S^{eff}_{RVB} = \int_{}^{}d\tau dx dy [a^\prime |\Psi|^2+b^\prime |\Psi|^4  - c^\prime |\Psi|^2 |\nabla \Psi|^2 ],
\end{eqnarray}
where $a^\prime<0$, $b^\prime>0$, $c^\prime<0$ are renormalized parameters
 from $a$,$b$,$c$ by coarse graining \cite{note1}.
 Note that the temporal phase stiffness ($\propto {1\over U}$, where $U$ is the local charging energy) is rather weak compared with the spatial phase stiffness $ E_s \propto |\Psi|^2$ and thus does not appear in Eq[\ref{EQ0}]. This is because charge fluctuation is significantly suppressed by strongly repulsive interactions, which leads to severe phase fluctuations thanks to the conjugating relation between phase variable and pair density. This accounts for the absence of phase coherence in half-filled RVB spin liquid. We note that the lack of phase coherence (or put it in a different way, the freezing of charge fluctuations) is what 
makes RVB spin liquid different from a superconductor of Cooper pairs, later on 
we will see how stripe dynamics helps to establish phase coherence in RVB spin liquid and turns it into a superconductor.

{\it{Hamiltonian of 1D stripes $H_{1D}$:}}
As shown by extensive experiments, stripes are dynamical in nature , and its dynamics includes the longitudinal charge and spin fluctuations along the stripe and
 the transverse motion which is relatively slow. Therefore it is valid to treat the stripes as 1D electron gas(1DEG) at first and then take the transverse degree of freedom into account by doing the suitable average over the transverse configurations. The general theory of Luttinger Liquid provides a powerful tool for the
 description of 1DEG.
 Following the notation of Luttinger Liquid \cite{LL}, the low energy effective theory of 1D stripes is described by separated charge modes and spin modes as follows \cite{caveat}:
\begin{eqnarray}
    H_{1D}^{eff}&=&\int_{}^{} dx [\frac{K_cu_c}{2}\Pi_c^2 + \frac{u_c}{2 K_c}{(\partial_x \Phi_c)}^2 ]            \\  \nonumber
          &+&\int_{}^{} dx [\frac{u_s}{2}\Pi_s^2+\frac{u_s}{2}{(\partial_x \Phi_s)}^2 + g_1 \cos(\sqrt{8\pi}\Phi_s) ] ,           
\end{eqnarray}
where $\Phi_c,~\Pi_c$,  and $ \Phi_s,~\Pi_s$  , are conjugated boson operators representing 
density fluctuations in charge and spin sectors of 1D Luttinger Liquid, respectively. $u_c$, $u_s$ are the corresponding propagating velocities, and $K_c$ is 
a parameter of interaction. The last term $g_1 \cos(\sqrt{8\pi}\Phi_s)$
 is the spin gap correction caused by the coupling between
 1D stripes and 2D RVB background with strong pairing correlations (see ref\cite{zheng} for detail).

{\it{Coupling between 1D stripes and 2D RVB background $H_{couple}$:}}
  
$ H_{couple}(c,c^+,d,d^+)=\sum _{k,q,\sigma}V c^+_{k,\sigma} d_{q,\sigma} \delta_{k_x,q} +h.c.  $,
where only horizontal stripe (along x direction) is considered, and momentum conservation is ensured by requiring
$k_x=q$. The coupling term accounts for the single electron hopping between 
stripes and the background and $V$ gives the hopping matrix element.  In order to discuss the low energy effective theory with boson variables only, we need to integrate out fermion variables so the lowest order relevant process happens at the second order of $V$. For the present discussion, we adopt the Cooper pair tunneling process \cite{EKT} as 
 the only effective coupling between 2D RVB environment and 1D stripes, that is 
\begin{eqnarray} 
  H^{eff}_{couple} (\Psi,\Delta)=g \int_{}^{} dx dy \Psi(x,y) \Delta^* (x) f(y-Y) +h.c. ,           \end{eqnarray}
where $g$ is the continuous limit of pair tunneling amplitude, $\Psi(x,y)$  stands for the coarse grained  RVB order parameter , $\Delta(x)$ is the singlet pairing order parameter of stripes, $f(y-Y)$ 
gives the transverse distribution function of the stripe position due to vibration , and  $Y$ stands for the transverse displacement of the stripe diffusion. 
 Here we assume that the stripe transverse modes can be described as the 
superposition of the fast mode of vibration  and the slow mode of diffusion.
Under the harmonic approximation, 
 one can assume $f(y)\propto \sqrt{\alpha}\exp(-\alpha y^2) $, where
 $1/\sqrt{\alpha}$ represents the amplitude of stripe vibration mode which 
 is determined by the microscopic details which are responsible for the stripe formation and stabilization. 
In Nd doped LSCO where stripes are pinned by the Low Temperature Tetragonal lattice structure \cite{stripe-exp},  $1/\sqrt{\alpha}$ is expected to be small 
(of the order of one lattice unit); while in optimum doped LSCO and YBCO, stripes are more disordered and transverse fluctuations are strong, so  $1/\sqrt{\alpha}$ can be as large as the order of inter-stripe distance which is about $4a$ (
$a$ is the lattice unit).

In the formulation of Luttinger Liquid \cite{LL}, $$\Delta(x)=d_u
\exp(-i\sqrt{2\pi}\Theta_c) \sin(\sqrt{2\pi}\Phi_s). $$

  We note that this pair tunneling process is very important in the present picture. Through this process the previously localized spin singlets in RVB background become mobile and the charge degree of freedom is resumed, in this sense one
 can no longer distinguish spin singlets of electrons from Cooper pairs which constitute the basis of superconductivity, and the RVB phase variable can be continuously connected to the phase degree of freedom of d-wave superconductivity order parameter. 

   After establishing the effective Hamiltonians of the coupled RVB and stripe variables, we can study the effect of stripe dynamics on the RVB background ( especially the phase degree of freedom ) by integrating out the stripe variables ( including both 
 the OP field  $\Delta(x)$ and the transverse mode variable $Y$)

\begin{eqnarray}
    \exp{[-\Delta S^{eff}_{RVB}]}&=&\int_{}^{} D\Phi_c(x,\tau)D\Phi_s(x,\tau) DY(\tau,x)  \\  \nonumber
  & &\exp[-\int_{0}^{\beta} d\tau (H_{1D}^{eff} + H^{eff}_{couple}+H_{tm})],
\end{eqnarray}
where $H_{tm}$is the Hamiltonian of the transverse modes of stripes,
$\beta={1\over{k_BT}}$.
Therefore, the integration up to the second order gives 
\begin{eqnarray}
\Delta S^{eff}_{RVB}&\approx&-\frac{g^2}{2}\int d\tau d\tau^\prime dx dy dx^\prime dy^\prime \Psi(x,y,\tau) \Psi^*(x^\prime,y^\prime,\tau^\prime)  \\   \nonumber
& &\langle\Delta^*(x,\tau)\Delta(x^\prime,\tau^\prime)\rangle_{1D} \langle f(y-Y(\tau,x))f(y^\prime-Y(\tau^\prime,x^\prime))\rangle_{tm}   
\end{eqnarray}
where $\langle~~~ \rangle_{1D}$ and $\langle~~~ \rangle_{tm}$ stand for average over longitudinal and transverse stripe variables, respectively.
According to the Luttinger Liquid theory,
\begin{displaymath}
\langle\Delta^*(x,\tau)\Delta(x^\prime,\tau^\prime)\rangle_{1D} \approx
\left\{ \begin{array}{ll}
\frac{d_u^2}{|(\tau-\tau^\prime)+i\frac{(x-x^\prime)}{u_c}|^{1\over{K_c}} |(\tau-\tau^\prime)+i\frac{(x-x^\prime)}{u_s}|} & |(\tau-\tau^\prime)+i\frac{(x-x^\prime)}{u_{s,c}}|<<{\xi_s\over{u_{s,c}}} 
\\
\frac{d_u^2}{|(\tau-\tau^\prime)+i\frac{(x-x^\prime)}{u_c}|^{1\over{K_c}} \xi_s/u_s} & |(\tau-\tau^\prime)+i\frac{(x-x^\prime)}{u_{s,c}}|>>{\xi_s\over{u_{s,c}}} 
\\
\end{array} \right.
\end{displaymath}
where $\xi_s\propto 1/\Delta_s$ is the cutoff of length scale given
 by the spin gap $ \Delta_s \propto   \sqrt{|g_1|} \exp{(\frac{v}{2\pi g_1})} $ \cite{zheng}.
 
   In order to extract the spatial and temporal phase stiffness coefficients, one can expand the integrand with respect to  $\Delta x=x^\prime - x$, $\Delta y=y^\prime - y$, $\Delta\tau=\tau^\prime - \tau$, over which one can perform integrations \cite{convergence}, then reach
\begin{eqnarray}
 \Delta S^{eff}_{RVB} \approx \int d\tau dx dy [E_\tau |\frac{\partial \Psi}{\partial \tau}|^2 +\Delta E_x |\frac{\partial \Psi}{\partial x} |^2 +\Delta E_y |\frac{\partial \Psi}{\partial y} |^2], \nonumber 
\end{eqnarray}
where the induced incipient temporal phase stiffness is
\begin{eqnarray}
 E_\tau &\propto& \delta g^2 \alpha \int_{0}^{\xi_s\over{u_{c,s}}} d\Delta\tau \langle \exp{[-\alpha(Y(\tau+\Delta\tau,x)-Y(\tau,x))^2/2]} \rangle_{tm} \\  \nonumber
& & (\Delta\tau)^{1-1/K_c}+\delta g^2 \alpha\int_{\xi_s\over{u_{c,s}}}^{\infty} d\Delta\tau \langle \exp{[-\alpha(Y(\tau+\Delta\tau,x)-Y(\tau,x))^2/2]} \rangle_{tm} \\  \nonumber
& & (\Delta\tau)^{2-1/K_c}, \nonumber
\end{eqnarray}

 Considering that $\langle (Y(\tau+\Delta\tau,x)-Y(\tau,x))^2 \rangle_{tm}=
2D(\Delta \tau)$ (where the stripe diffusion is modeled as random walk and $D$ is the diffusive coefficient), then the above expression can be simplified to \cite{note2}
\begin{eqnarray}
\label{EQ1}
 E_\tau \propto \delta g^2 \alpha (\alpha D)^{1/K_c-2} F(\frac{\alpha D \xi_s}{u_c})
 +\frac{u_{c,s}}{\xi_s}\delta g^2 \alpha (\alpha D)^{1/K_c-3} G(\frac{\alpha D \xi_s}{u_c}),
\end{eqnarray} 
where $ F(X)=\int_{0}^{X} x^{1-1/K_c}e^{-x} dx $,
$G(X)=\int_{X}^{\infty} x^{2-1/K_c}e^{-x} dx$.
At the limit $\frac{\alpha D\xi_s}{u_c}  >>1$, $E_\tau \propto \delta g^2 \alpha (\alpha D)^{1/K_c-2}$; while for $\frac{\alpha D\xi_s}{u_c} <<1$, $E_\tau \propto \delta g^2\alpha (\alpha D)^{1/K_c-3}/\xi_s$. 

The corrections to spatial phase stiffness $\Delta E_x $ and $\Delta E_y$ can be calculated similarly, however, compared with the unperturbed $E_s$ of 2D RVB effective theory in eq[\ref{EQ0}]
 they are negligibly small when $\delta$ is small enough.

  Now combined with eq[\ref{EQ0}] where the spatial phase stiffness $E_s\propto \rho^2$ is given, and retain only phase variables (assuming frozen amplitude $|\Psi|=\rho$), one can discuss the phase ordering process in the RVB liquid with the following effective action:
\begin{eqnarray}
S_{eff}&=&\int_{0}^{\beta}d\tau [\sum_{i} E_\tau\rho^2 |\frac{\partial \phi(\vec r_i)}{\partial \tau}|^2  \\ \nonumber 
&-& E_s\rho^2 \sum_{<ij>} \cos(\phi(\vec r_i)-\phi(\vec r_j))]. \nonumber 
\end{eqnarray} 
Notice here we turn from the Ginzburg-Landau like "soft-spin" effective model into "hard spin" XY model, because in (2+1) dimension they belong to the same universal class and thus have the same critical behavior.

This effective action has been under heavy discussions in the study of granular superconducting film and Josephson junction array\cite{JJ}. To reveal the 
quantum critical physics inside it, one can perform a standard Hubbard-Stratonovich transformation to decouple the Josephson term \cite{doniach}, which introduces the complex order-parameter field $\psi$ in proportion to the expectation value of $\exp(i\phi)$. The resulting Ginzburg-Laudau action in (2+1)D reads:
\begin{eqnarray}
F_{eff}[\psi]&=&\int dxdyd\tau [\frac{1}{8E_s\rho^2}|\nabla \psi|^2+128E_\tau^3\rho^6|\frac{\partial\psi}{\partial\tau}|^2  \\ \nonumber  
 &+& ({1\over{2E_s\rho^2}}-4E_\tau\rho^2)|\psi|^2+\kappa |\psi|^4 ].
\end{eqnarray}
So the quantum critical point(QCP) is given by ${1\over{2E_s\rho^2}}-4E_\tau\rho^2=0$, which separates the zero-temperature phase diagram into superconducting ordered phase ($E_\tau E_s \rho^4>1/8$)
 and non-superconducting disordered phase
($E_\tau E_s \rho^4<1/8$). At finite temperature, there exists a crossover temperature  $T_{cr}(\rho,E_\tau) \propto \sqrt{\frac{|E_\tau\rho^2-\frac{1}{8E_s\rho^2}|}{E_\tau^3 \rho^6}}$, above which lies the quantum critical region where physical quantities obey scaling laws with $T$. On the SC ordered side the crossover temperature becomes the transition temperature corresponding to the well-known KT transition \cite{KT}. The phase diagram is shown in Fig.1(a). Note that strong asymmetry exists in $T_{cr}$ around the QCP, and the much higher crossover temperature on the disordered 
 side compared with the SC ordered side can explain why the anomalous $T$ dependent scaling behaviors are prevalent in the normal states of superconducting cuprates while in 
slightly doped insulating cuprates the critical regime eludes experiments (it is however likely that stripe ordering itself can lead to critical scaling behavior which is not considered here). 

\section {Discussions and Conclusions}
Now let us discuss how to connect this QCP with the general phase diagram of high $T_c$ cuprates. One can see the quantum phase transition is tuned by 
a single parameter $E_\tau E_s \rho^4 \propto \delta g^2 \rho^6 H(\xi_s,\alpha, D)$
which is a complicated function of doping density $\delta$, RVB OP amplitude $\rho$, spin gap $1/\xi_s$
 and stripe transverse mode parameters $\alpha$ and $ D$. In realistic experiments, upon hole doping, all the other parameters change accordingly. For example, RVB OP amplitude $\rho$ and spin gap both decrease with doping (experimentally  spin gap closes around $\delta=0.2$), while stripe transverse modes may depend on material-dependent properties like lattice distortions and impurity effects. Therefore a comprehensive understanding of this issue can be formidable and will not be pursued here. However For the 
purpose of qualitative demonstration 
of the physical mechanism , I will attempt to take some of the relevant parameters into account (while leave the others like those of stripe
 transverse modes as external inputs)  and mark the route 
 followed by a cuprate under hole doping in the ground state phase diagram (Fig.1(b)).At first, with 
slightly doping from the parent cuprate, $E_\tau$ increases from zero (roughly in proportion to $\delta$)
 while $\rho$ gradually decreases from the maximum, therefore at a critical doping value ($\delta_{c1}\approx 0.06$) the system crosses the phase boundary into the SC ordered state. Then upon further doping from under-doped to over-doped regions, $\delta$ gradually becomes saturated, meanwhile a diminishing spin gap pushes $E_\tau$ toward the limit value controlled by $\alpha$ and $D$. Therefore the route is bent toward $E_\tau$ axis thanks to the decreasing $\rho$ (because over-doping reduces RVB correlations significantly with excessive holes "overflowing" into the background, which is also consistent with the result of RVB mean field calculations \cite{laterRVB} ). Finally as $\delta>\delta_{c2}\approx 0.3$  the system crosses the phase boundary again and returns to the disordered non-superconducting ground state. During the above process, $T_{cr}$ increases from zero to its maximum and then decreases back to zero, as it is the case for the transition temperature \cite{note3}.

  Before end, two comments are in order. First, I will comment on the role of transverse stripe modes in 
affecting the SC transition. According to eq[\ref{EQ1}], lower $\alpha$ 
 and $D$ tends to strengthen $E_\tau$ (which is especially effective in under doped region where the spin gap is substantial and the limit $\frac{\alpha D\xi_s}{u_c} <<1$ can be approached, assuming $K_c>1/2$ which coincides with the condition under which SC fluctuations along stripes are relevant at low energy \cite{stripephase}). This suggests that in the present 
mechanism, larger transverse vibration amplitude ($\approx 1/\sqrt{\alpha}$)
 favors SC while the diffusion mode does not. Considering the various 
 stripe phases proposed in literature \cite{stripephase}, it is interesting 
to note that SC order is favored only in the intermediate region between the stripe crystal phase ( with small vibration amplitude, or large $\alpha$ )
 and stripe liquid phase ( where stripes are meandering strings and diffusions dominate), which implies a very subtle relation between 
SC order and stripe charge order.  
Secondly, I will briefly compare the present picture with the one suggested in 
 ref \cite{EKT}: in that work, the superconducting order is induced by the Josephson tunneling between neighboring stripes and it is natural to expect this coupling to be strongly dependent on the inter-stripe distance (presumably decays exponentially with the distance ) and also the extent of disorder in stripe configurations, which makes it a subtle issue to explain 
the simple and well-defined relation between $T_c$, zero temperature superfluid density and doping density, and the fact that higher $T_c$ is found in the cuprates with more disordered stripe correlations. In the present work, the induced 
temporal phase stiffness only depends on the local coupling between one stripe and its neighboring background and is therefore not sensitive to the disorder in the coupling between the neighboring stripes.

   In conclusion, the low energy effective theory of the RVB phase variable coupled to the stripe dynamics is obtained, where the effect of stripe dynamics induces doping dependent incipient temporal phase stiffness in the RVB liquid, which tunes a quantum phase transition toward a superconducting ground state with global phase order.

   I am grateful to S.A.Kivelson for discussions . The support from Stanford Graduate Fellowship (SGF) and SSRL is gratefully acknowledged.

%%%%%% Fig 1 %%%%%%
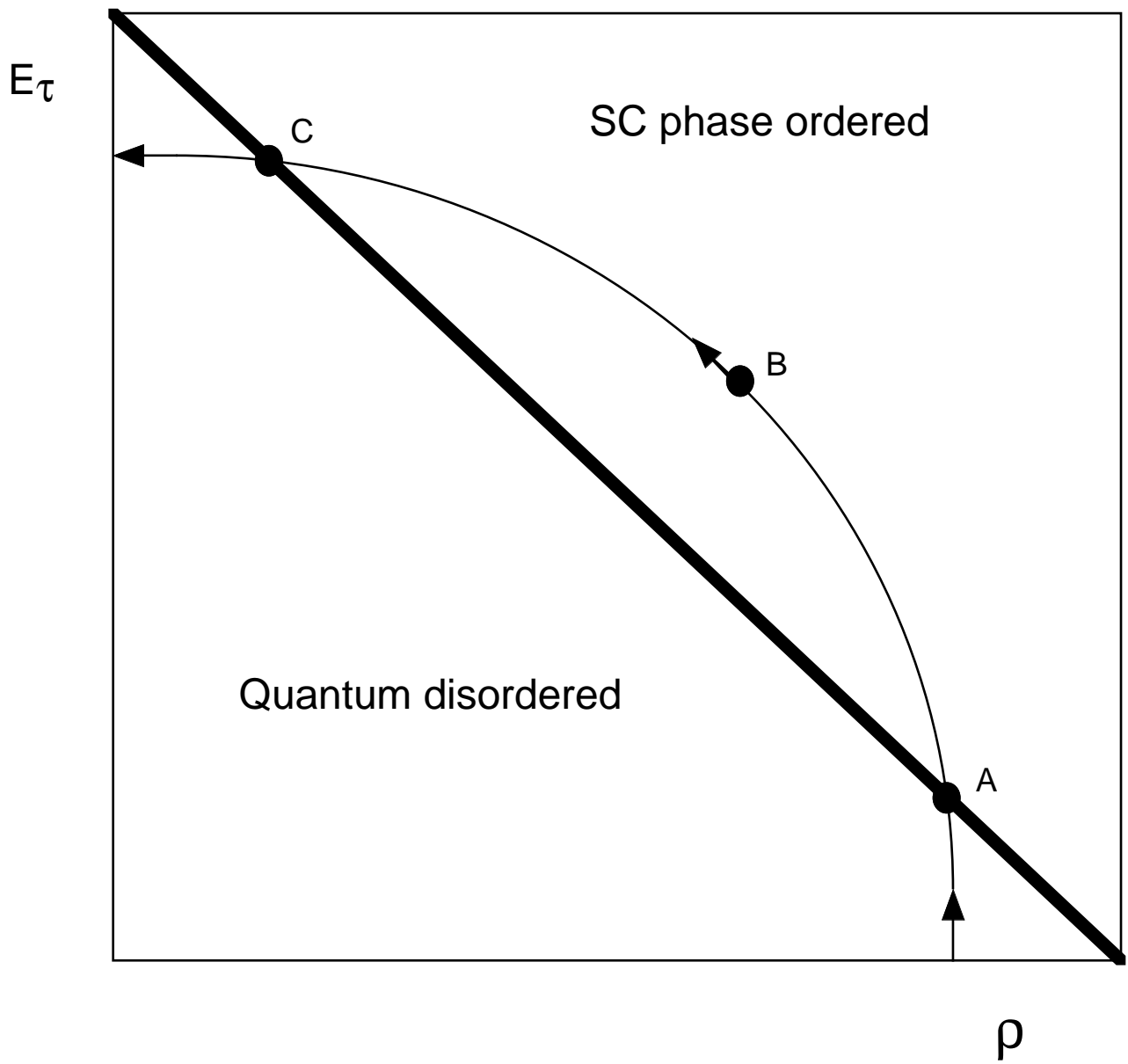
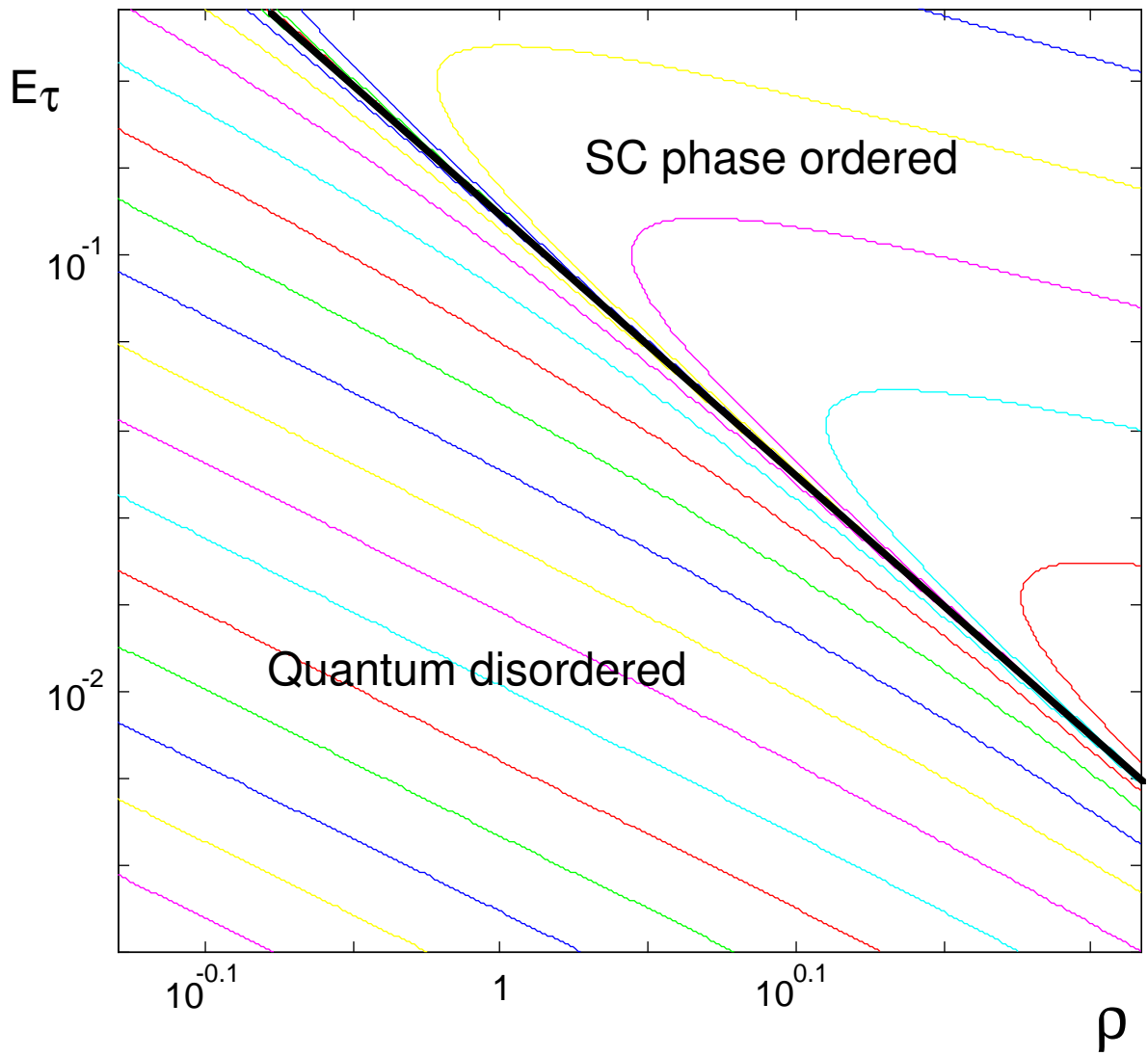
\begin{figure}
\caption{
(a).The zero temperature phase diagram of the effective action describing RVB phase variable coupled to stripe dynamics. The thick line marks the phase boundary between the SC ordered and disordered 
 phases, while the thin curves represent contours with equal $T_{cr}$'s.
(b).The route followed by a cuprate under hole doping in the ground state phase diagram. $A \rightarrow B$ and $B \rightarrow C$ correspond to under-doped and over-doped regions, respectively. Both axes use logarithmic scale with arbitrary unit.
}
\end{figure}


\begin{references}
\bibitem{SDW} D.Pines, Physica C{\bf 282-287}, 273 (1997);
S. Sachdev and J. Ye, Phys. Rev. Lett.{\bf 69}, 2411 (1992).  
\bibitem{zhang} S. C. Zhang, Science {\bf 275}, 1089 (1997). 
\bibitem{laughlin} R. B. Laughlin, cond-mat/9709195.
\bibitem{stripe-exp} J. M. Tranquada et al., Nature (London) {\bf 375}, 561 (1995); K. Yamada et al., Phys. Rev. B{\bf 57}, 6165(1998); G. Aeppli et al., Science {\bf 178}, 1432 (1997); A. W. Hunt et al., Phys. Rev. Lett.{\bf 82}, 4300 (1999); Z. X. Shen et al., Science {\bf 280}, 259 (1998).
\bibitem{italy} C. Castellani et al., Phys. Rev. Lett.{\bf 75}, 4650 (1995); J. Zaanen, cond-mat/9811078.
\bibitem{stripesc}A. H. Castro Neto, Phys. Rev. Lett.{\bf 78}, 3931 (1997);  M. Granath and H. Johannesson, Phys. Rev. Lett.{\bf 83}, 199 (1999);
S. R. White and D. J. Scalapino, Phys. Rev. B.{\bf 60}, 753 (1999).
\bibitem{EKT} V. J. Emery , S. A. Kivelson, and O. Zachar, Phys. Rev. B{\bf 56}, 6120 (1997).
\bibitem{EK1} S. A. Kivelson et al.,cond-mat/9707327.
\bibitem{zheng} W. J. Zheng, Phys. Rev. Lett.{\bf 83},3534(1999).
\bibitem{note0} It is noted that the present scenario 
  is in fact not limited to the RVB picture, instead, it only relies on the existence of preformed but incoherent pairs with strongly suppressed 
 temporal phase stiffness.   
\bibitem{anderson} P. W. Anderson, Science {\bf 64}, 188 (1986).
\bibitem{laterRVB} D. H. Lee, cond-mat/9909111; P. A. Lee, cond-mat/9812226, and references therein.
\bibitem{EK2} V. J. Emery and S. A. Kivelson, Nature {\bf 374},434 (1995).
\bibitem{anderson2} G. Baskaran and P. W. Anderson, Phys. Rev. B{\bf 37}, 580(1988).
\bibitem{note1} $a^\prime<0$ is ensured at $T<<J$ ($J$ is the AFM exchange coupling), and $b^\prime>0 $ is required for stability against amplitude fluctuations, as $c^\prime<0$ is for the stability 
against phase fluctuations. Here we limit ourselves to the consideration of uniform SC ground state, excluding the possibilities of anomalous SC orders at finite momentum or with self-generated flux. 
\bibitem{LL}For a review, see " Bosoniztion and Strongly Correlated Systems", by A. O. Gogolin, A. A. Nersesyan, and A. M. Tsvelik (Cambridge University Press, 1998).
\bibitem{caveat} There is however proposal that 1D Luttinger Liquid theory may not be relevant to stripe physics, see O. Tchernyshyov, L. P. Pryadko, cond-mat/9907472 and S. R. White (private communication).
%\bibitem{fnote} 
\bibitem{convergence} The integrations over $\Delta \tau$ and $\Delta x$ are convergent under the assumption of transverse stripe diffusion along $\tau$ and $x$ axis which gives exponential decaying $<ff>_{tm}$. 
Therefore the gradient expansion is valid here. 
\bibitem{note2} Note that doping density $\delta$ is introduced by the spatial average over inter-stripe distance, which is proportional to $1/\delta$ for $\delta<1/8$. For $\delta>1/8$, $\delta$ is replaced by its saturated value $1/8$.
\bibitem{JJ} M. P. A. Fisher et al., Phys. Rev. B{\bf 40}, 546(1989);
S. Chakravarty et al., Phys. Rev. B{\bf 37}, 3283(1988). 
\bibitem{doniach} S. Doniach, Phys. Rev. B{\bf 24}, 5063(1981).
\bibitem{KT} J. M. Kosterlitz and D. J. Thouless, J. Phys. C {\bf 6},1181 (1973).
%\bibitem{pin} J. M. Tranquada et al., Phys. Rev. B{\bf 54}, 7489(1996); K. %Hirota et al.,physica B {\bf 241/243}, 817(1998). 
\bibitem{note3} Despite the qualitatively similar doping evolution, the transition temperature is quantitatively different from $T_{cr}$ in that the former corresponds to truly 3D ordering and depends on factors more than the intra-plane correlations (which however contains the essential physics for superconductivity), such as inter-plane couplings and characteristics of charge reservoir layers etc.
\bibitem{stripephase} H. Eskes et al., cond-mat/9712316; S. A. Kivelson, 
E. Fradkin and V. J. Emery, Nature {\bf 393}. 550 (1998).
\end{references}
\end{document}